\documentclass[12pt]{article}
\usepackage{amssymb,amsmath,graphicx}

\begin{document}

\begin{titlepage}

\begin{center}

\hfill ICRR-Report-523-2005-6\\
\hfill KEK-TH-1055\\
\hfill \today

\vspace{0.8cm}

{\large Efficient Coannihilation Process through Strong Higgs Self-Coupling
in LKP Dark Matter Annihilation}

\vspace{1.3cm}

{\bf Shigeki Matsumoto}$^{a,\,}$\footnote{smatsu@post.kek.jp}
and 
{\bf Masato Senami}$^{b,\,}$\footnote{senami@icrr.u-tokyo.ac.jp} \\

\vspace{1cm}

{\it
$^a${Theory Group, KEK, Oho 1-1, Tsukuba, Ibaraki 305-0801, Japan}\\
$^b${ICRR, University of Tokyo, Kashiwa 277-8582, Japan }
}

\vspace{1.5cm}

\abstract{
We point out that the lightest Kaluza-Klein particle (LKP) dark matter
in universal extra dimension (UED) models
efficiently annihilates % into the standard model particles
through the coannihilation process including the first KK Higgs bosons
when the Higgs mass is slightly heavy as $200 - 230$ GeV,
which gives the large Higgs self-coupling.
The large self-coupling naturally leads
the mass degeneracy between the LKP and the first KK Higgs bosons
and large annihilation cross sections of the KK Higgs bosons.
These are essential for the enhancement of the annihilation of the LKP dark matter,
which allows large compactification scale $\sim 1$ TeV to be
consistent with cosmological observations for the relic abundance of dark matter.
We found that the thermal relic abundance of the LKP dark matter could be reconciled with
the stringent constraint of electroweak precision measurements in the minimal UED model.
}

\end{center}
\end{titlepage}

%%%%%%%%%%%%%%%%%%%%%
%%%%%%%%%%%%%%%%%%%%%
\section{Introduction}
%%%%%%%%%%%%%%%%%%%%%
%%%%%%%%%%%%%%%%%%%%%

There are some compelling evidence to require an extension of the standard
model (SM), for example, the existence of non-baryonic cold dark matter,
neutrino masses and the baryon number asymmetry in the universe. Many models
beyond the standard model are proposed to solve one or a few of these problems.
In particular, extensions to explain the existence of dark
matter are promising, because many direct and indirect detection experiments
for dark matter are now on going, and many future experiments are also proposed.
Therefore the extensions will be tested in near future.

Many models including a dark matter candidate have been proposed.
Among those, models with a weakly interacting massive particle (WIMP) are attractive.
This is because the WIMP can naturally provide the correct relic abundance of
dark matter in the present universe in addition to the successful explanation
of the large scale structure of the universe.

A famous candidate for WIMP is the lightest supersymmetric particle (LSP)
in a supersymmetric extension of the SM \cite{susydm}, which is stabilized by the
R-parity. Recently, the lightest Kaluza-Klein particle (LKP) in the flat
universal extra dimension (UED) scenario \cite{Appelquist:2000nn} has been
proposed as an alternative candidate for WIMP. In UED models, all particles
in the SM propagate in the compact extra dimensions. The momentum along
an extra dimension is interpreted as the Kaluza-Klein (KK) mass in the four
dimensional point of view. The KK mass spectrum is quantized in terms of KK
number $n$ as $m^{(n)} =  n/R$, where $R$ is the size of the extra dimension.
The momentum conservation along the extra dimension leads to the KK
number conservation. Since UED models must contain the SM as a low energy
effective theory, extra dimensions are compactified on an orbifold for
deriving chiral fermions. This orbifolding breaks the KK number conservation
down to the KK-parity conservation,
in which the SM particles and even KK particles carry $+1$ charge
while odd KK particles carry $-1$ charge.
Due to the KK-parity conservation,
the LKP is stable and a good candidate for dark matter.

The thermal relic abundance of the LKP dark matter is calculated in several
papers \cite{Servant:2002aq, Kakizaki:2005en, Kong:2005hn, Burnell:2005hm}
and the results indicate the compactification scale $1/R$ consistent with
observations such as WMAP \cite{WMAP} is in the range of 500 to 700 GeV in
the minimal UED (MUED) model. The MUED model is defined in the five
space-time dimensions in which the extra dimension is compactified on an
$S^1/Z_2$ orbifold. The one-loop corrected KK mass spectrum of the model
are calculated in Ref.~\cite{Cheng:2002iz}.

On the other hand, the scale $1/R$ is constrained by electroweak
precision measurements (EWPM) \cite{Appelquist:2000nn,Appelquist:2002wb,EWPM,Flacke:2005hb}.
The constraints in most of the previous papers \cite{Appelquist:2000nn,Appelquist:2002wb,EWPM}
are satisfied for $1/R \gtrsim$ 300 GeV,
however, a severe constraint is recently
reported as $ 1/R >$ 700 GeV (for $m_h$ = 120 GeV) at the $99$\% confidence
level \cite{Flacke:2005hb}. This result seems to indicate that the MUED model
is inconsistent with the observed abundance of dark matter.

However, we found a parameter region 
reconciling the relic abundance of dark matter with
the stringent constraint reported in Ref.~\cite{Flacke:2005hb}
at the 3$\sigma$ level in the MUED model.
The parameter region is where the Higgs mass is slightly heavy as $m_h \gtrsim$ 200 GeV.
In the region,
the self-coupling of the SM Higgs field (and its KK particles) becomes large.
Therefore the annihilation cross sections of the KK Higgs particles are enhanced.
Furthermore, the first KK particles of the Higgs field are degenerated with the LKP
(the first KK photon) in mass, which is less than $1\%$ level in the MUED model.
As a result, the thermal relic abundance of the LKP is significantly changed
through the coannihilation process including the first KK Higgs bosons.
To be more precise, the first KK charged and pseudo Higgs bosons are
important for the coannihilation process,
which are the first KK particles of the SM Goldstone bosons.
Since the first KK particle of the neutral scalar Higgs is heavier
as the SM Higgs is heavier,
it does not contribute significantly to the coannihilation process.

This paper is organized as follows.
In the next section, we briefly review the MUED model,
especially we focus on the masses of the LKP and the first KK Higgs bosons.
Next we discuss the thermal relic abundance of the LKP dark matter
in the section \ref{abundance}.
We evaluate the compactification scale consistent with observations of the relic abundance
depending the value of the SM Higgs boson mass.
Section \ref{conclusion} is devoted to summary and discussion.

%%%%%%%%%%%%%%%%%%%%%
%%%%%%%%%%%%%%%%%%%%%
\section{Mass difference between LKP and first KK Higgs bosons}
%%%%%%%%%%%%%%%%%%%%%
%%%%%%%%%%%%%%%%%%%%%

The simplest version of UED model i.e. MUED model has one extra dimension
compactified on an $S^1/Z_2$ orbifold.
Thus the model is described by the SM with extra particle contents
which are the KK modes of the SM particles in the four-dimensional point of view.
The SM particles and their KK particles have identical charges
and couplings relevant for KK particles are completely determined by those of SM.
Hence, the MUED model has only two new input parameters,
the compactification scale $1/R$ and the cutoff scale $\Lambda$.
The cutoff scale is usually taken to be
$\Lambda R \sim {\cal O}(10)$ \cite{Appelquist:2000nn}.
In this paper, we adopt the value, $\Lambda R = 20$,
since the changing this value only slightly modifies our results.
In addition to these parameters,
we have one undetermined parameter,
that is the mass of the SM Higgs boson, $m_h$.
This parameter is very important for our studies
because the masses of the KK Higgs bosons and
the self-coupling between them depend on this value.

The mass spectra of KK particles are determined by $1/R$ and
the mass of the corresponding SM particle at tree level.
One of the typical features of the MUED model is
that all particles at each KK level are degenerated in mass.
Thus, the radiative corrections to the masses play an important role
when we consider the mass difference between KK particles \cite{Cheng:2002iz}.
Below, we summarize the masses including the radiative
corrections for the first KK photon and first KK Higgs bosons,
which are important for our discussion.

The LKP is the first KK photon in most of the parameter region in the MUED model.
Its mass is obtained by diagonalizing the mass squared matrix
described in the $(B^{(1)}, W^{3(1)})$ basis, 
\begin{eqnarray}
  \left(
  \begin{array}{cc}
   1/R^2 + \delta m^2_{B^{(1)}} + g^{\prime 2} v^2/4 & g'g v^2/4 \\
   g'g v^2/4 & 1/R^2 + \delta m^2_{W^{(1)}} + g^2 v^2/4
  \end{array}
 \right) ,
 \label{eq:LKP_mass_matrix}
\end{eqnarray}
where $g(g')$ is the SU(2)$_L$(U(1)$_Y$) gauge coupling constant
and $v \simeq 246$ GeV is the vacuum expectation value of the SM Higgs field. 
The radiative corrections to the massive KK gauge bosons are given by
\begin{eqnarray}
\delta m^2_{B^{(1)}}
 &=&
 -\frac{39}{2}\frac{g^{\prime 2}\zeta(3)}{16\pi^4 R^2}
 -\frac{1}{6}\frac{g^{\prime 2}}{16\pi^2 R^2}
  \ln \left(\Lambda^2 R^2\right) , \\
\delta m^2_{W^{(1)}} &=&
 -\frac{5}{2}\frac{g^2\zeta(3)}{16\pi^4 R^2}
 +\frac{15}{2}\frac{g^2}{16\pi^2 R^2}
  \ln \left(\Lambda^2 R^2\right) .
\end{eqnarray}
The difference between diagonal elements, $\delta m^2_{W^{(1)}} -
\delta m^2_{B^{(1)}}$, exceeds the off-diagonal ones when $1/R \gg v$.
Hence, the weak mixing angle of the first KK gauge bosons is small
and the KK photon is dominantly composed by
the first KK particle of the hypercharge gauge boson.

Let us turn to the masses of the first KK Higgs particles. 
Since the KK modes of Higgs field are not eaten by the SM gauge bosons,
all of neutral scalar $H^{(1)}$, pseudoscalar $A^{(1)}$ and
charged scalar $H^{\pm(1)}$ remain as physical states.
The latter three states are the KK particles of the Goldstone modes in the SM.
The KK Higgs boson masses turn out to be
\begin{eqnarray}
 m_{H^{(1)}}^2 &=& 1/R^2 + m_h^2 + \delta m_{H^{(1)}}^2 ,
 \label{eq:higgsmass} \\
 m_{H^{\pm (1)}}^2 &=& 1/R^2 + m_W^2 + \delta m_{H^{(1)}}^2 ,
 \label{eq:chargedhiggsmass} \\
 m_{A^{(1)}}^2 &=& 1/R^2 + m_Z^2 + \delta m_{H^{(1)}}^2 ,
 \label{eq:pseudohiggsmass}
\end{eqnarray}
where $m_W$ and $m_Z$ are $W$ and $Z$ boson masses, respectively.
The radiative correction $\delta m_{H^{(1)}}^2$ is given by
%%%%%%%%%%%%%%%%%%%%%
\begin{eqnarray}
 \delta m_{H^{(1)}}^2 = \left( \frac{3}{2}g^2 + \frac{3}{4}g^{\prime 2} - \lambda_h \right)
 \frac{1}{16\pi^2 R^2} \ln \left(\Lambda^2 R^2\right) ,
 \label{eq:deltamh}
\end{eqnarray}
where $\lambda_h$ is the Higgs self-coupling defined as
%%%%%%%%%%%%%%%%%%%%%
%\begin{eqnarray}
$\lambda_h \equiv m_h^2/v^2 .$
%\end{eqnarray}
%%%%%%%%%%%%%%%%%%%%%
As increasing $m_h$, $\lambda_h$ becomes large
and the negative contribution in Eq.~(\ref{eq:deltamh}) increases. 
Hence, for large $\lambda_h$,
the annihilation cross sections of the KK Higgs bosons are significantly enhanced 
and the mass differences between the LKP and $H^{\pm(1)}(A^{(1)})$ become small. 
However, the mass differences are negative when $m_h$ is too large.
Thus the LKP becomes the charged KK Higgs boson
and this case is not allowed from the point of view of dark matter
\footnote{We oversighted the charged LKP region in the previous papers
\cite{Kakizaki:2005en} and noticed it recently.}.

%%%%%%%%%%%%%%%%%%%%
\begin{figure}[t]
\begin{center}
\scalebox{.9}{\includegraphics*{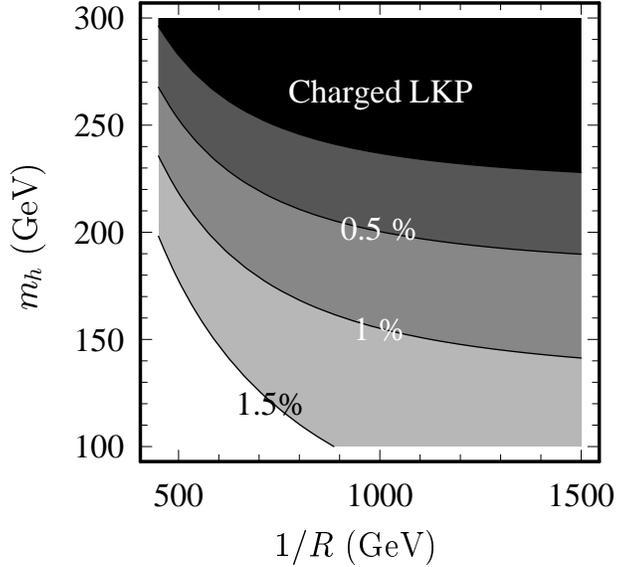}}
\caption{\small The contour map of the mass degeneracy between the first KK photon
   and charged KK Higgs, $(m_{H^{(1)\pm}} - m_{\gamma^{(1)}} )/ m_{\gamma^{(1)}}$
   in ($1/R$, $m_h$) plane. The cutoff scale $\Lambda$ is set to be $\Lambda R =20$.}
\label{fig:lkpcont}
\end{center}
\end{figure}
%%%%%%%%%%%%%%%%%%%%

In Fig.~\ref{fig:lkpcont}, we depict the mass degeneracy
between the LKP and the charged KK Higgs boson, 
$(m_{H^{\pm(1)}} -  m_{\gamma^{(1)}})/ m_{\gamma^{(1)}}$.
It is clear form this figure 
that the mass difference between the first KK charged Higgs boson and the first KK photon
is very small for $m_h \sim 150 - 230$ GeV.
On the other hand, 
$H^{\pm(1)}$ is identified with the LKP for $m_h \gtrsim 250$ GeV, 
so that this parameter region should be discarded from our discussion. 

Here, we should address the KK particle of the graviton.
Since a radiative correction to the mass of the KK graviton
is extremely small due to the gravitational interaction,
the mass is given by $1/R$ with high accuracy.
Furthermore, the value, $( m_{\gamma^{(1)}} - 1 / R ) $,
is positive for $1/R \lesssim 800 $ GeV,
the LKP is the first KK graviton in this region.
This fact leads us to a serious problem, that is,
the graviton LKP dark matter with mass less than O(10) TeV is excluded
due to the late time decay of the NLKP (next LKP) to the KK graviton
and it is severely constrained by cosmological observation for cosmic 
microwave background \cite{Feng:2003xh}.

Fortunately, the region we are interested in is $1 /R \gtrsim 800$ GeV
for $ m_h \sim 220 -230 $ GeV, in which the relic abundance of dark
matter is consistent with cosmological observations.
In the region the KK graviton is not the LKP,
and the problem discussed above is replaced with
the problem caused by the late time decay of the KK graviton to the LKP.
This is avoided if the reheating temperature of the universe is low enough 
\cite{gravitino}.

Furthermore, there is the mechanism that makes only the KK graviton
become heavy without changing other sectors \cite{Dienes:2001wu}.
The mechanism is based on higher dimensional setup than
that used in the MUED model,
and KK particles in the MUED are assumed to be localized in
the five dimensional space-time.
With the use of the mechanism,
the LKP dark matter is identified with the KK photon in the MUED model
even if $1/R$ is less than 800 GeV.

%%%%%%%%%%%%%%%%%%%%%
%%%%%%%%%%%%%%%%%%%%%
\section{Relic abundance of the LKP dark matter revisited}
\label{abundance}
%%%%%%%%%%%%%%%%%%%%%
%%%%%%%%%%%%%%%%%%%%%

We are now in a position to calculate the thermal relic abundance of the LKP
dark matter including the coannihilation with the first KK Higgs bosons, 
especially when the SM Higgs is slightly heavy. 
Since the LKP is also degenerate with the first KK leptons in mass, 
the coannihilation 
processes including these particles should be taken into consideration as well. 
For the detailed formulae of the mass spectra, refer to Ref.~\cite{Cheng:2002iz}.

We use the method developed in Ref.~\cite{Griest:1990kh}
to calculate the relic abundance including the coannihilation effects.
Under reasonable assumptions,
the relic density of the LKP obeys the following Boltzmann equation,
\begin{eqnarray}
 \frac{dY}{dx} =
 -\frac{\langle \sigma_{\rm eff} v \rangle}{Hx}s \left( Y^2 - Y_{\rm eq}^2 \right) ,
 \label{eq:Boltzmann}
\end{eqnarray}
where $Y = n/s$ and $x = m/T$.
The number density $n$ is defined as the sum of the number density of each species $i$
as $n \equiv \sum_i n_i$.
The entropy density is given by $s = (2 \pi^2 / 45) g_*m^3/x^3$, 
with $g_* = 86.25$ being the relativistic degrees of freedom at the decoupling. 
The Hubble parameter is $H = 1.66g_*^{1/2}m^2/x^2 m_{\rm Pl}$,
where $m_{\rm Pl} = 1.22\times 10^{19}$ GeV is the Planck mass.
The abundance in the equilibrium $Y_{\rm eq}$ is written as
\begin{eqnarray}
 Y_{\rm eq} = \frac{45}{2\pi^4} \left(\frac{\pi}{8}\right)^{1/2}
 \frac{g_{\rm eff}}{g_{*}} x^{3/2} e^{-x} ,
\end{eqnarray}
where $g_{\rm eff}$ is the number of the effective degrees of freedom
and defined by
\begin{eqnarray}
 g_{\rm eff} = \sum_i g_i(1 + \Delta_i)^{3/2} e^{-x \Delta_i} ,
 \qquad
 \Delta_i = (m_i - m_{\gamma^{(1)}})/m_{\gamma^{(1)}} .
\end{eqnarray}
The number of the internal degrees of freedom for species $i$ is denoted by $g_i$.

The effective annihilation cross section $\sigma_{\rm eff}$ 
governs the relic density of the LKP dark matter and
is given as the sum of $\sigma_{ij}$,
which denotes the coannihilation cross section between species $i$ and $j$,
\begin{eqnarray}
 \sigma_{\rm eff} = \sum_{i,j} \sigma_{ij} \frac{g_i g_j}{g_{\rm eff}^2} 
 (1 + \Delta_i)^{3/2} (1 + \Delta_j)^{3/2} \exp[-x(\Delta_i + \Delta_j)] .
 \label{effective CS}
\end{eqnarray}
The annihilation cross section, $\sigma_{ij}$,
in each process has been already calculated.
For the explicit expressions,
see Refs.~\cite{Servant:2002aq,Kong:2005hn,Burnell:2005hm}. 
By solving the Boltzmann equation numerically, 
we obtain the present abundance of dark matter, $Y_\infty$.
It is useful to express the relic density in terms of $\Omega h^2 = m n h^2/\rho_c$,
which is the ratio of the dark matter density to the critical density
$\rho_c = 1.1 \times 10^{-5}\ h^2\ {\rm cm}^{-3}$. 
The small letter $h$ denotes the scaled Hubble parameter, $h = 0.71^{+0.04}_{-0.03}$.

%%%%%%%%%%%%%%%%%%%%
\begin{figure}[t]
\begin{center}
\scalebox{1.3}{\includegraphics*{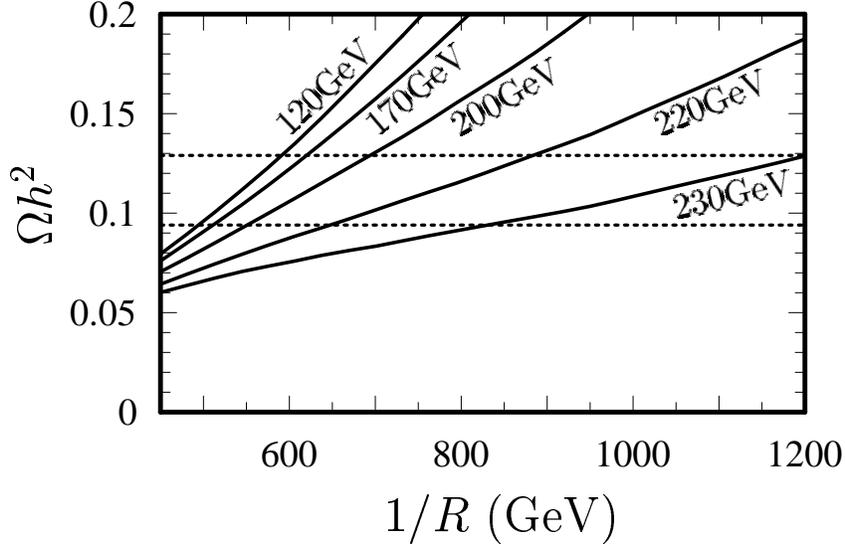}}
\caption{\small Thermal relic abundance of the LKP dark matter as a function of $1/R$.
  The solid lines correspond to the results with
  $m_h = 120, 170, 200, 220$, and $230$ GeV from top to bottom.
  Two horizontal dashed lines are the allowed region from the WMAP observation 
  at the 2$\sigma$ level, $0.094 < \Omega h^2 < 0.129 $.}
\label{fig:abundance}
\end{center}
\end{figure}
%%%%%%%%%%%%%%%%%%%%

The results of the calculation are shown in Fig.~\ref{fig:abundance}.
The thermal relic abundance of the LKP dark matter is depicted as a function of
$1/R$ with the SM Higgs mass, $m_h = 120, 170, 200, 220$, and $230$ GeV.
Two horizontal dashed lines denote the allowed region from the WMAP
observation at the 2$\sigma$ level, $0.094 < \Omega h^2 < 0.129 $ \cite{WMAP}.

It is found that the larger compactification scale is allowed for the larger Higgs mass.
The is because the large Higgs self-coupling is derived for the larger Higgs mass.
The large Higgs self-coupling induces two effects.
First, the mass differences between $\gamma^{(1)}$ and $H^{\pm(1)} (A^{(1)})$
become small for the larger Higgs self-coupling.
Therefore the Boltzmann suppression factor $(1 + \Delta_i)^{3/2}\exp(-x\Delta_i)$
in the effective annihilation cross section in Eq.~(\ref{effective CS}) become negligible.
Second, the annihilation cross sections
between the first KK Higgs bosons are significantly enhanced.
As a result, the effective annihilation cross section also increases,
and the compactification scale as large as $1/R \sim 1$ TeV
can be consistent with the observed relic abundance of dark matter.

%%%%%%%%%%%%%%%%%%%%%
%%%%%%%%%%%%%%%%%%%%%
\section{Summary and discussion}
\label{conclusion}
%%%%%%%%%%%%%%%%%%%%%
%%%%%%%%%%%%%%%%%%%%%

In this paper, we have investigated the dependence of the relic abundance of
the LKP dark matter on the SM Higgs mass in the MUED model.
It is found that the effective annihilation cross section governing the
abundance is drastically enhanced when $m_h \sim 200 - 230$ GeV
and the compactification scale consistent with the observed abundance increases.
The key ingredient is the strong Higgs self-coupling, 
which allows the LKP dark matter to annihilate very efficiently
in the early universe through the coannihilation processes
including the first KK Higgs bosons.
Due to the enhancement of the coannihilation processes for $m_h \gtrsim 200$ GeV,
the relic abundance consistent with cosmological observations could be produced
without conflicting the bound reported by Ref.~\cite{Flacke:2005hb}
at the 3$\sigma$ level.

Finally, we address the implication of the parameter region
we investigated to the Large Hadron Collider (LHC), which starts at CERN in $2007$.
In order to satisfy both bounds from
the abundance and the EWPM reported by Ref.~\cite{Flacke:2005hb},
a slightly heavy Higgs is required.
Since large mass of the light Higgs scalar in the minimal supersymmetric model
is not favored, it may be a signature of the UED models
if we observe a heavy Higgs boson and missing momentums at the LHC.

%%%%%%%%%%%%%%%%%%%%%
\section*{Acknowledgements}
%%%%%%%%%%%%%%%%%%%%%
We are grateful to M. Kakizaki for useful discussions.
The work of SM was supported in paart by a
Grant-in-Aid of the Ministry of Education,
Culture, Sports, Science, and Technology,
Government of Japan, No. 16081211.

%%%%%%%%%%%%%%%%%%%%%
%%%%%%%%%%%%%%%%%%%%%

\end{document}